\documentclass[10pt,conference]{IEEEtran}

\usepackage{graphicx}
\usepackage{subfig}
\usepackage{siunitx,booktabs}
\usepackage{listings}
\usepackage{color}
\usepackage{multirow}
\usepackage{bigstrut}
\usepackage{url}
\usepackage{array}
\usepackage{ctable}
\usepackage{amsmath}
\usepackage{flushend}

\newcommand{\code}[1]{{\small\textsf{#1}}}
\newcommand{\etal}{~\textit{et al}.~}

\author{\IEEEauthorblockN{Phong Minh Vu, Tam The Nguyen, Tung Thanh Nguyen}

\IEEEauthorblockA{Department of Computer Science and Software Engineering\\
Auburn University\\
\{lenniel,tam,tung\}@auburn.edu}\\
}

\begin{document}
\title{On building an automated responding system for app reviews: What are the characteristics of reviews and their responses?}
\maketitle

\begin{abstract}
Recent studies showed that the dialogs between app developers and app users on app stores are important to increase user satisfaction and app's overall ratings. However, the large volume of reviews and the limitation of resources discourage app developers from engaging with customers through this channel. One solution to this problem is to develop an Automated Responding System for developers to respond to app reviews in a manner that is most similar to a human response. Toward designing such system, we have conducted an empirical study of the characteristics of mobile apps' reviews and their human-written responses. We found that an app reviews can have multiple fragments at sentence level with different topics and intentions. Similarly, a response also can be divided into multiple fragments with unique intentions to answer certain parts of their review (e.g., complaints, requests, or information seeking). We have also identified several characteristics of review (rating, topics, intentions, quantitative text feature) that can be used to rank review by their priority of need for response. In addition, we identified the degree of re-usability of past responses is based on their context (single app, apps of the same category, and their common features). Last but not least, a responses can be reused in another review if some parts of it can be replaced by a placeholder that is either a named-entity or a hyperlink. Based on those findings, we discuss the implications of developing an Automated Responding System to help mobile apps' developers write the responses for users reviews more effectively.

\end{abstract}
\section{Introduction}

Mobile app development is increasingly competitive with millions of apps are currently available on popular app markets. Therefore, improving the user satisfaction of mobile apps is of importance to their developers. Recent studies suggest that responding to users' reviews of an app could increase its overall ratings~\cite{mcilroy2017worth}. However, few user reviews are actually responded (i.e. less than 1\% reviews get response~\cite{mcilroy2017worth}), possibly due to the huge volumes of reviews and the lack of resources from the development teams. Furthermore, many users may write reviews about similar things~\cite{vu2015MARK} (e.g. a topic, an issue, a suggestion, etc). If developers answered to one of those reviews, the rest should also be answered similarly. Therefore, there is a need to develop a system that can suggest to developers which review to answer based on their past answers. Such a system would reduce the workload for developers and increase their efficiency in dealing with user reviews. To make this system possible, it is crucial to answering the research question: What are the characteristics of the responded reviews? Moreover, the system should be also able to leverage the past responses to help developers to build new ones for similar reviews. Thus, we also need to answer the research question: What are the characteristics of the review responses? 


Two previous studies~\cite{hassan2017reviewDialog, mcilroy2017worth} have initially explored several aspects of responses for app reviews, including the dialogues of between developers and users and how ratings affected responses. They have several important findings that motivated our works, such as: (1) Developer responses have a positive impact on user rating; (2) Responses can be repetitive, which means an Automated Response System can re-use past responses; (3) Different apps may have different likelihood of a developer responding; (4) Several quantitative metrics of reviews can have impact on the likelihood of getting a responses (e.g. rating, text length, sentiment, title); (5) There are several drivers for responding (e.g. endorsement, thanking, advertising, asking for detail, updates).

\begin{figure*}
\sf
\scriptsize
\centerline{\includegraphics[width=0.8\textwidth]{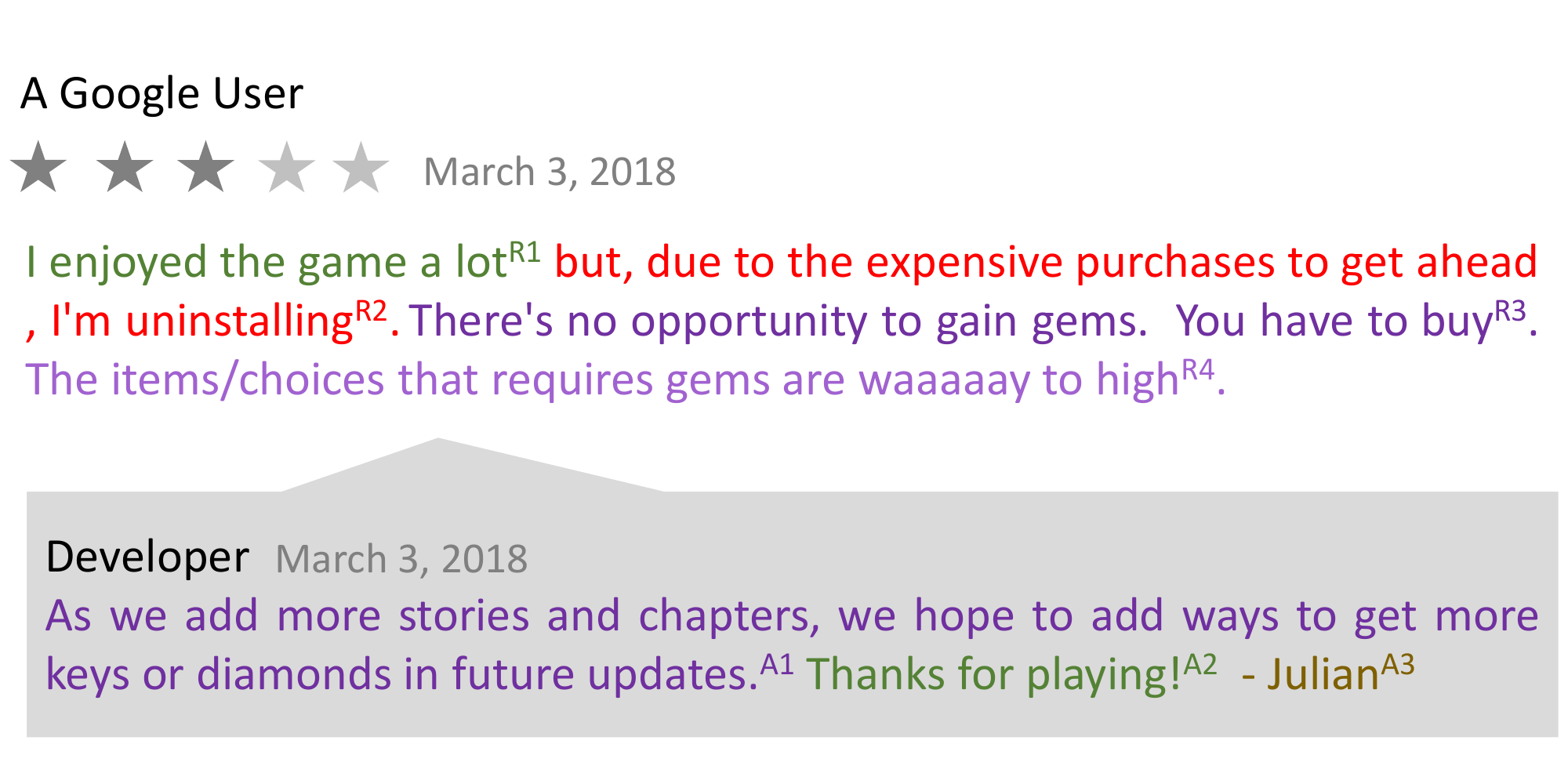}}

\caption{A pair of review and response}
	\label{fig:example}
\end{figure*}

However, for the purposes of creating an Automated Responding System, there are several more specific research questions have not been answered:
\begin{itemize}
    \item \textbf{What are the components of reviews and how do they affect the likelihood of getting a response?} Previous studies~\cite{surf,reviewresponse} suggested that reviews can have multiple topics and intentions. We believe they could have an effect on the response as well. However, the above empirical studies did not cover this part.
    \item \textbf{What are the components of responses and how do they make up a response as a whole?} We believe that the seven drivers of responding are not detailed enough to generate each response, as our observation on our data have pointed out a response can be composed by multiple sentences with different intentions (Table \ref{tab:egResponse}). 
    \item \textbf{What are the characteristics of re-usable responses?} Hassan\etal found that some responses can be re-used, and some are not meant to be. However, there is yet a complete analysis of what response/parts of the response to re-use, and to what extent.
    \item \textbf{What are the structure of a full response and how to generate it?} An Automated Responding System needs to be able to generate a suitable response for the review of interest. To do this, we need to understand what a generation model can create for the response structure before filling it with content.
\end{itemize}

To answer those questions, we have conducted an empirical study on the characteristics of reviews and their responses. In this empirical study, we collected and manually labeled 3,212 pairs of review and response from 33 top trending apps on Google Play store for three first months of 2019. Furthermore, we used this dataset to answer the aforementioned questions and were able to discover several important characteristics of responses and their reviews. One of such characteristics is the segmented nature of reviews and responses. For example, the pair of review-response in Figure \ref{fig:example}, is annotated into different segments. The review's segments are:  R1 is a Praise; R2 is a Complaint about how expensive the purchases are; R3: is a Complaint about how users have no other option than buying gems; R4 is a Complaint about the content that needed gem are expensive. Similarly, the response's segments are: A1 is a Promise about adding more ways to earn gems and keys in the future (targeted to complain R2, R3, and R4); A2 is an Appreciation; A3 is the developer's name. Further observations can be found in Table \ref{tab:egReviews} and \ref{tab:egResponse}, which gives the intention, topic, and re-usability of review and response segments.

The example above is perhaps not an uncommon sight in reviews and their responses. However, this particular review has very specific and informative complaints, and the response is composed of different parts that served different purposes for those complaints. In further observations, we found similar reviews without a response that could have been responded in the same manner, such as: ``\textit{Great game, the amount of diamonds it costs to unlock special choices compared with how much they cost and the ability to gain them massively let's the game down sadly. Be prepared to miss out a lot of the options unless you spend major amounts of money}''. This may indicate that we can re-use responses for similar reviews to an extent. Further investigation had discovered that the re-usability of a response may depend on the review's topics and intentions. Moreover, we also found that, while the category of apps can affect the re-usability of responses, Google Play Store's categories may not be reliable for that purpose. In addition to that, not all responses can be reused as-is, but rather be re-used in the form of canned-reply, or template. We discovered that the average templated responses have 1.6 placeholders (e.g. named entity, hyperlinks).

Last but not least, our findings also suggest that a Markov Chain model is a feasible way to generate the components of a response based on its review.

The rest of this paper is as following: Section \ref{sec:data} describes our data and the annotation process. Section \ref{sec:reviews} studies the characteristics of responded reviews. Section \ref{sec:response} studies the characteristics for the responses. Section \ref{sec:discussion} discusses about how the findings from our studies affect the future Automated Responding System. Section \ref{sec:validity} discusses about the construction, internal, and external threats to the valididy of our research. Next, Section \ref{sec:related} discusses about all the related work to us. Finally, Section \ref{sec:conclusion} concludes our paper.
\section{Data collection}
\label{sec:data}
In this section, we describe our approach for collecting the user-developer responses from the Google Play store and how we labeled the data.

\subsection{Apps selection}

To select suitable apps for the study, we followed several criteria:

\begin{itemize}
    \item Top apps from different categories on Google Play store: We expect the top apps to have an adequate amount of user-base, which would lead to a high amount of user reviews.
    \item App has to be matured: App needs to have at least 6 months of history on Google Play store.
\end{itemize}

We ran our crawler from January 1, 2019,  to  March 6, 2019. During that period, we collect 649,645 reviews for 164 apps. More than 3000 collected  reviews  have  received  a  response  from  the  app  developer as shown in Table \ref{tab:dataset}. The percentage of replied reviews is about 0.5\% in this dataset.

\begin{table}[htbp]
  \centering
  \caption{Overview of our dataset}
   
    \begin{tabular}{|l|r|}
    \hline
    App number & 164 \bigstrut\\
    \hline
    Number of reviews & 649,645 \bigstrut\\
    \hline
    Time period & Jan 1, 2019 to Mar 6, 2019 \bigstrut\\
    \hline
    Apps with response & 33 \bigstrut\\
    \hline
    Number of Responses & 3212 \bigstrut\\
    \hline
    \end{tabular}%
  \label{tab:dataset}%
\end{table}%

\subsection{Data annotation for the studies}
 
 \subsubsection{Coding method}
To fully answer the research questions that will be subsequently discussed in the following sections, we labeled our data of 3212 reviews and their responses. 

The previous studies suggested that intentions and topics can describe an opinion, or an idea \cite{alpaca,surf}. Topics can be described by keywords, while intentions can be classified by linguistic patterns at the sentence and phrase levels. However, separating sentences into phrases would require too much labor, hence, we decided to annotate data at the sentence level by their intentions and topics. We used ALPACA's~\cite{alpaca} text normalizer module to split the 3,212 pairs of reviews and responses into 18,301 sentences and used them as input for our annotation process. 

On another matter, a prior study has discussed the possibility of re-using the past responses on newer reviews. This posed the need to distinguish which responses and which sentences are re-usable as well. Therefore, we also marked the sentences that are deemed re-applicable for a different review/app.
 
To annotate intentions, we followed these three steps:
\begin{itemize}
    \item Step  1: our  authors  independently  used the Open Coding Method \cite{strauss1997grounded,khandkar2009open} to identify the intentions in sentences. This method  iteratively  builds  a  list  of  identified  intentions.  For every studied sentence, the author identifies its intentions started with 5 basic intentions mentioned by the authors of SURF~\cite{surf}. If the intention is not in the list of identified intentions, the author extends the list and revisits all sentences using the new list of identified intentions. It terminates when there are no newer intentions to be identified and all sentences are studied. Note that some sentence can have more than one intention. We discusses the intentions in the next subsection.
    \item Step  2: The  first  author  compared  the  intentions  that  were  identified  by  all authors for all studied sentences and marked the differences.
    \item Step 3: The authors discussed the differences and came to a consensus about the  final  intentions  for  each  studied  sentence.  We  found  15  sentence intentions that  had  differences  in  the  identified  intentions  between  the  two  coders.  After discussion, all differences were resolved. The inter-rater agreement was 99\% (almost perfect agreement), measured  by the Cohens  kappa  coefficient~\cite{fleiss1973equivalence}.
\end{itemize}

For topics, we used the 12 general topics from a previous study by Hassan\etal. The coders followed a similar three steps to the Open Coding Method, but on the first step, the coders just choose one of the topics and not create any new one. In case a sentence cannot be assigned a topic (e.g. \code{``a bit annoying'' does not belong to any topic}, we simply ignored it. After the discussion, the inter-rater agreement was 99\%  measured by the Cohens Kappa coefficient. 

Similarly, we marked the re-usable sentences with the same method. Our inter-rate agreement rate was also 99\%.

For other smaller-scale studies in this paper, we also applied the same process of labeling. 

\begin{table}[htbp]
  \centering
  \caption{The labeled data's overall statistic}
   
    \begin{tabular}{|l|r|}
    \hline
    Number of sentences & 18,127 \bigstrut\\
    \hline
    Number of review sentences & 7,095 \bigstrut\\
    \hline
    Number of response sentences & 11,032 \bigstrut\\
    \hline
    Number of review intention & 8 \bigstrut\\
    \hline
    Number of response intention & 11 \bigstrut\\
    \hline
    \end{tabular}%
  \label{tab:dataset}%
\end{table}%

\begin{table*}
\small
  \centering
  \caption{Examples of Review sentences and their annotations}
    \begin{tabular}{p{30em}p{8em}p{8em}}
    \toprule
    \textbf{Review sentence} & \textbf{Topic} & \textbf{Intention}  \\
    \midrule
    the latest update turns the screen black and blank after page load & GUI, Update/Version & complaint \\
    \midrule
    there's no ability sort yourself by language & feature  & complaint \\
    \midrule
    facing error when downlaod any file after minimize the app or turn off screen the downloading will stop it shows retrying-retrying & download, feature & complaint \\
    \midrule
    nice app for every social media solution & app, feature & praise \\
    \midrule
    remove minimum order value 200 for shopping coupan & pricing & request \\
    \midrule
    4444.00 & no topic & unknown \\
    \bottomrule
    \end{tabular}%
  \label{tab:egReviews}%
\end{table*}%

\begin{table*}
\small
  \centering
  \caption{Examples of Response sentences and their annotations}
    \begin{tabular}{p{30em}p{8em}p{8em}}
    \toprule
    \textbf{Review sentence} & \textbf{Intention} & \textbf{Re-usable}  \\
    \midrule
    hey justin  & greeting & yes \\
    \midrule
    badoo is free to use but for those of you who want to stand out from the crowd and interact with our top rated users we also offer a large range of premium services such as credits and super powers  & information giving  & no \\
    \midrule
    our sync article may help with connecting your calendar to outlook goo.gl/bmfd5c & solution & no \\
    \midrule
    also you can follow the instructions here to submit this feedback directly to our development team https://goo.gl/5ebqty  & customer support & yes \\
    \midrule
    we're so sorry to hear about this vinit  & apology & yes \\
    \midrule
    thanks for your feedback  & appreciation & yes \\
    \bottomrule
    \end{tabular}%
  \label{tab:egResponse}%
\end{table*}%
\subsection{Review Intentions Labels}

In this sub-section, we briefly discuss each of the intentions defined in our annotation process for reviews.
\subsubsection{Comparison} comparing with other apps/products/platforms/developers. Example: \code{``operations on some sites are not good such as filling applications uploading photos are not good as compared to google chrome''}
   
\subsubsection{Complaint} any negativity remark (i.e. problem description, denouncement, bug report, negative emotional speech). Example: \code{``the game had an issue with false advertising of there ingame merchandise''}.
   
\subsubsection{Request} any request (i.e. feature request, asking for freebies, asking for help, suggestion, etc). Example: \code{``please its like two months i am being asking and requesting you for blood bound book two ''}.
   
\subsubsection{Information Giving} giving any information that is not bug description, or may not related to the app. The intention is to inform other users or developers. Example: \code{``i am connected to the internet''}.
   
\subsubsection{Information seeking} asks for information or suggests that user need certain information in return (even without the question mark or any question). Example: \code{``plzz tell me how to download movie from your app in android mobile''}.
   
\subsubsection{Praise} praises, appreciates explicitly. Example: \code{``keep the excellent job''}.
   
\subsubsection{Ultimatum} uses leverage to ask for something in return or just threatens the developers without asking for anything. Example: \code{``my rating will return to five stars when the problem is fixed''}.
   
\subsubsection{Unknown} no clear intention. Example: \code{``so we'll see''}.

\subsection{Response Intentions Labels}
In this sub-section, we briefly discuss each of the intentions defined in our annotation process for responses.
\subsubsection{Solution} offering a solution to a problem. Example: \code{``as a first step please try clearing your browsing data using the instructions in our chrome help center https://goo.gl/nijk5e''}.
   
\subsubsection{Customer support} offering customer support. Example: \code{``if you can't go live in imo could you please send your imo account with country code and your problem to us''}.
   
\subsubsection{Greeting} any form of greeting. Example: \code{``hey nazzer''}.

\subsubsection{Promise} giving promises of a specific problem/suggestion. Example: \code{``we ll get this looked at for you''}.
   
\subsubsection{Appreciation} telling the customer how grateful the developers are. Example: \code{``thank you for your review''}.

\subsubsection{Apology} apologizing to user. Example: \code{``we re sorry to hear of your experience''}.
   
\subsubsection{Farewell/signing name} goodbye/giving the responder's name. Example: \code{``have a great day''}.
   
\subsubsection{Information seeking} asks for more information, or opinions. Example: \code{``would you be able to provide us with more details on what went wrong or could you perhaps amend the review if you chose the low rating by accident''}.
   
\subsubsection{Asking for re-rating} asking for a re-rating of the app. Example: \code{``in the meantime we'd appreciate it if you would consider giving us a 5 star rating on google play as this rating is very important to us in building trust with our users''}.
   
\subsubsection{Information giving} giving information that belongs to neither of the above intentions. Example: \code{``you can read more about this in our terms and conditions''}.
   
\subsubsection{Unknown} no clear intention. Example: \code{``that doesn't sound good''}.


\section{Study of the characteristic of responded reviews}
\label{sec:reviews}
A popular app can have more review than developers can respond, resulting in the low percentages of responded reviews~\cite{mcilroy2017worth}. Moreover, app stores may limit the number of daily responses from developers (e.g. Google Playstore has a limit of 500 responses a day). Therefore, we believe it is necessary to determine what kind of reviews the developer should prioritize responding to. To answer this question, in this section, we explore the characteristics of reviews that have been responded in our dataset. Please note that, even though Hassan\etal's work has described some aspects related to this study, we further expand it to the statistics of intentions and sentences for a more complete overview of the matter.
\subsection{Dataset}
We used the truth set of 3,212 reviews that have been labeled for this study, as it contains all responded reviews in the whole dataset. While similar statistics can be obtained from the rest of the dataset, there was no indicator that all of them have been read and chosen to not be responded by the developer. Thus, the statistics of non-responded review is not a reliable source for this study.

\subsection{How does rating affect the likelihood of being responded?}
\begin{figure}
 \caption{Percentage of each rating in replied reviews}
	\includegraphics[width=0.45\textwidth]{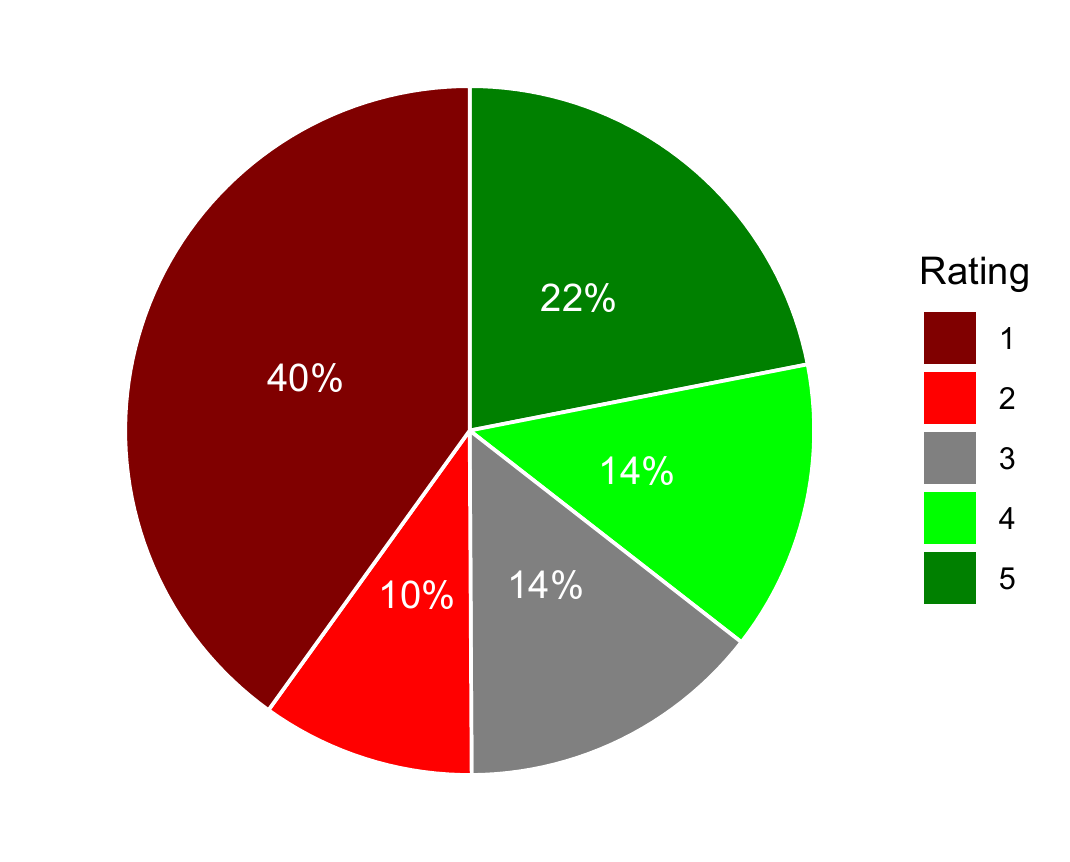}
 	\label{fig:ratingPie}
\end{figure}
As shown in Figure \ref{fig:ratingPie}, half of the responded reviews have a rating lower than 3. These numbers further reinforce the finding that developers are likely to respond to lower rating reviews from the previous study~\cite{hassan2017reviewDialog}. However, the five-star ratings ranked second in this dataset. This \textbf{suggests that rating alone may not be a sufficient feature to indicate if a review is worth responding or not}.

\subsection{What are the topics of responded reviews?}
\begin{table}[t]
  \centering
  \caption{General topics mentioned in responded reviews}
    \begin{tabular}{lr}
    \toprule
    \textbf{Topics} & \textbf{Percentage} \\
    \hline
    Feature/Functionality & 52\% \\
    App & 38\% \\ 
    Content & 38\% \\
    Update/version  & 30\% \\
    GUI  & 27\% \\
    Improvement & 18\% \\
    Pricing & 10\% \\
    Resources & 8\% \\
    Security  & 4\% \\
    Company  & 3\% \\
    Model & 1\% \\
    \bottomrule
    \end{tabular}%
  \label{tab:topics}%
\end{table}%

To answer this question, we classified our review sentences using SURF's 12 general topics. The results are shown in Table \ref{tab:topics}. The top topics in this dataset are Feature/Functionality, Update/Version, GUI, Improvement, App, and Content. This \textbf{suggests that some reviews topics are more prone to get a response than others.}

\subsection{What are the intentions of responded reviews?}
\begin{table}[t]
  \centering
  \caption{Intentions in responded reviews}
    \begin{tabular}{lr}
    \toprule
    \textbf{Intentions} & \textbf{Percentage} \\
    \hline
    complaint &  47\% \\
    praise &  39\% \\
    request &  37\% \\
    Information giving &  24\% \\
    Information seeking & 14\% \\
    Ultimatum/threatening &  7\% \\
    comparison &  6\% \\
    rating &  5\% \\
    unknown & 2\% \\
    \bottomrule
    \end{tabular}%
  \label{tab:purposes}%
\end{table}%

Table \ref{tab:purposes} showed our results of how intentions are distributed in the responded review sentences. We observed that some interesting intentions such as praise, request, complaint, information giving are dominant in this data. This \textbf{suggests that praise, request, complaint, and information giving are the specific intentions that developers of these apps were looking for to answer}.

\subsection{How are the other quantitative text features?}

We suspected that some quantitative text features (number of sentences, number of words, number of intentions, number of topics) may also affect the decision of developers to respond to a review. Therefore, we collected the statistics of those text features from our responded reviews and showed it in Table \ref{tab:quantitative}. While this quantitative data does not explain fully why they are responded, it may \textbf{suggest that reviews with at least 2 topics and intentions are likely to be responded.}

\begin{table}[t]
  \centering
  \caption{Some quantitative Text Features of Responded reviews}
    \begin{tabular}{lr}
    \toprule
    \textbf{Features} & \textbf{} \\
    \hline
    Average words count &  24.6 \\
    Average sentence count &  2.2 \\
    Average intention count & 1.8 \\
    Average topic count &  2.3 \\
    \bottomrule
    \end{tabular}%
  \label{tab:quantitative}%
\end{table}%
\section{Study of the characteristics of responses}
\label{sec:response}
For our automated system to construct a response, we have to understand the structure, content, and sentiment that a response is supposed to deliver. Therefore, in this section, we conducted a study of the responses in our labeled dataset on two aspects: the intentions, the re-usability of past responses, and the templates of re-usable responses.

\subsection{What are the intentions in responses?}
\begin{table}
  \centering
  \caption{Intentions in responses}
    \begin{tabular}{lr}
    \toprule
    \textbf{intentions} &  \textbf{Percentage} \\
    \hline
    Appreciation &  32\% \\
    Greeting &  30\% \\
    Information giving &  24\% \\
    Customer Support &  19\% \\
    Promise &  12\% \\
    Apology &  10\% \\
    Solution &  6\% \\
    Information seeking & 5\% \\
    Asking for re-rating & 4\% \\
    Farewell/Signing & 4\% \\
    unknown & 4\% \\
    \bottomrule
    \end{tabular}%
  \label{tab:resIntent}%
\end{table}%

As shown in Table \ref{tab:resIntent}, there are 11 intentions that can be found in response. The most common intentions are appreciation and greeting, which we believe are common courtesy to improve the relationship with the customers. The next prominent intentions include Information Giving, Customer Support, Promise, Apologies, and Solution. As we read through the responses, we have noticed that sentences containing those intentions are usually short and re-usable, with the exception of Solutions and Information Giving, as shown in Table \ref{tab:egResponse}. The Customer Support sentences usually contain a link to an email or external website. We believe the reason for this phenomenon might be of the following: (1) developers cannot disclose customer's private information on the public response; (2) the problem might need further communication between the customer support team and the customer. An email thread/customer support system is a better choice to keep track of them. 

\subsection{To what extent can we re-use past responses?}

In our dataset, we have found that 81\% of responses sentences were generic enough to be re-used in another review for a similar effect. However, simply placing those responses into a new review is not possible, as they may be referring to a different company, app name, customer name, etc, in their text. In this sub-section, we seek to answer the question: to what extent can we re-use past responses.

\subsubsection{Re-usability of past responses in single app}

In the context of a single app, it is common for the same problem can occur to multiple users. In such cases, it should be easier for the developer to just copy and paste the content of previous responses to answer a similar review. For example, the sentence \code{``as a first step please try clearing your browsing data using the instructions in our chrome help center https://goo.gl/nijk5e''} for Google Chrome app offers a solution to complaints about screen turned black by customers and was actually re-used \textbf{58 times} in our dataset for Google Chrome. We suspected that the reason why these reviews are answered similarly is because of their similar specific topics and intentions, as they together formed an opinion\cite{vu2015MARK}. 

\begin{table}
  \centering
  \caption{Overview of 11 apps that have more than 100 responded reviews}
    \begin{tabular}{lr}
    \toprule
    \textbf{Intentions} &  \textbf{Number of reviews} \\
    \hline
    UC Browser& 	350 \\
    Uber& 	337 \\
    MX Player& 	335 \\
    Google Chrome& 	254 \\
    PicsArt Photo Studio& 	217 \\
    Last Day on Earth& 	177 \\
    Robinhood& 	130 \\
    Google Translate& 	129 \\
    Tinder& 	121 \\
    imo free video calls and chat& 	120 \\
    Google Calendar& 	104 \\

    \bottomrule
    \end{tabular}%
  \label{tab:11Apps}%
\end{table}%

To validate this theory, we grouped the reviews by their topics and intention for 11 apps (Table \ref{tab:11Apps}) that has more than 100 responded reviews as these apps have 2,274 reviews, or 70.8\% of the total. After that, we manually check if the reviews in the group had similar or the same responses and compute the percentage of similar responses. Table \ref{tab:topicSimilar} shows the results from the topics/intentions with at least 50 reviews in the analyzed set. Overall, almost all topics have a high degree of re-usability for their responses, however, this was largely due to the highly generic responses that can fit in any review. For example, in MX Player, the majority of responses for any feature, or functionality complaint would be \code{``Hi,Seems like you are facing some issues on our App. Please write to us at support@j2apps.com with your issues in detail and we will assist you with the same. Meanwhile, hit 5 star and show your support for MX.''}, which does not address the problem directly, but rather redirect it to Customer Support. Similarly, for intention,  it seems that the Praise intention always gets a scripted appreciation response, while more important ones such as Complaint and Request would sometimes get a specific response. The only exception is Information Seeking reviews, which mostly get an answer that addresses the question.

\textbf{Conclusion: Some topics and intentions are likely to be responded by a re-used response.}

 \begin{table}[htbp]
   \centering
   \caption{Percentage of similar responses for each topic and intention}
     \begin{tabular}{lrlr}
     \toprule
     \textbf{Topic} & \multicolumn{1}{l}{\textbf{}} & \multicolumn{1}{l}{\textbf{Intention}} & \multicolumn{1}{l}{\textbf{}}\\
     \midrule
    Security  & 100\% & Praise & 100\%\\ 
    Company  & 100\%  & Information Giving & 85\%\\
    Improvement & 100\% & Complaint & 71\%\\
    Feature/Functionality & 91\% & Request & 68\%\\
    App & 87\% & Information Seeking & 25\%\\ 
    GUI  & 82\% &  \\
    Content & 79\% & & \\
    Pricing & 78\% & & \\
    Update/version  & 62\% & & \\
     \bottomrule
     \end{tabular}%
   \label{tab:topicSimilar}%
 \end{table}%

\subsubsection{Re-usability of the past responses in multiple apps}

Based on the labeled data, there is a strong indication that past responses can be used across multiple apps. However, it is safe to assume that the re-usability of those responses is not the same as for a single app, as each app can have a different context, with its own set of problems and features. Only some generic types of responses are re-used. For example, the sentence used by the Google Chrome team in the previous sub-section cannot be reused for other apps because they clearing browsing data and Chrome Help Center are two unique features to Chrome. In contrast, the sentence \code{``screenshots would be helpful''} can be used for any bug report in any app. Moreover, a previous study~\cite{Chen:2014:AR-miner} suggested that the similarity of app reviews can vary depending on the groups of app (i.e. all apps versus apps of the same category). Therefore, it is necessary to study the re-usability of past responses based on each app category.
\begin{table}
  \centering
  \caption{Categorization using Google Play Store on our dataset}
    \begin{tabular}{lr}
    \toprule
    \textbf{Categories} &  \textbf{Number of Apps} \\
    \hline
    Communication& 7 \\
    Finance&	6 \\
    Game&	3 \\
    Tools&	3 \\
    Photography&	2 \\
    Personalization&	2 \\
    Entertainment&	2 \\
    Social&	2 \\
    Maps and Navigation&	1 \\
    Lifestyles&	1 \\
    Productivity&	1 \\
    Music and Audio&	1 \\
    News and Magazines&	1 \\
    Shopping&	1  \\
    \bottomrule
    \end{tabular}%
  \label{tab:categories}%
\end{table}%

To further investigate what kind of reviews can be re-used, we use the categories in the Google Play store split our 33 apps into 14 categories. In this study, we only analyze the categories with at least 3 apps to avoid unfair comparison: Communication, Finance, Games, and Tools. Within the limited scope of this study, we took 100 random reviews from each category and manually analyze if their response can be used for other apps. The results are shown in Table \ref{tab:ReusableCategories}. As we can see, Finance responses have 100\% re-usability, while Games' responses have 68\%. In our observation of the data, Finance apps can re-use all of their responses because they did not attempt to solve any problem directly by responding, instead, they referred to an external Customer Support system for a more private and personal solution to their customers. This policy is consistent with all 6 apps. Similarly to Games, if the response was to redirect to Customer Support, then it can be re-used. Other responses are only re-usable if they did not mention the specific game feature asked in the review. Lastly, Communication and Tools apps are mostly very different apps in nature, thus, their responses are often not reusable by other apps in the same category. However, we also found that they may be re-usable for apps that share some same features with them. For example, Google Chrome has responses that redirect the reader to their external forums. This is not popular within its category (i.e. Communication), but it is quite popular in the Game category, as all the games have an external forum for their players to discuss. This suggests that Google Play Store's categorization of apps may not be a sufficient source to categorize apps in the reusable aspect of reviews.

\textbf{Conclusion: Google Play Store's categories may not be reliable to categorize re-usability of app review responses.}
\begin{table}
  \centering
  \caption{Percentage of reusable reviews within each category}
    \begin{tabular}{lr}
    \toprule
    \textbf{Categories} &  \textbf{Re-usable reviews} \\
    \hline
    Finance&	100\% \\
    Game&	68\% \\
    Communication& 17\% \\
    Tools&	0\% \\
    \bottomrule
    \end{tabular}%
  \label{tab:ReusableCategories}%
\end{table}%

\subsection{What are response sentence templates?}

During the labeling process, we have observed that many responses were actually canned-responses, which means, they have a template. For example, the sentence \code{``Let us know by sending a quick note to t.uber.com/contact and we'll connect''} can be used for any app with a customer support link if we replace \code{``t.uber.com/contact''} with that link. In further observations, we found that there are several types of placeholders like that, such as [user name], [company name], [link], and more. By replacing the appropriate placeholders with relevant text, we can have a similar response that fits in with another review. To further investigate what are the templates of responses, we randomly select 100 sentences from the dataset that were labeled to be reusable, then, we identify the parts that can be replaced and re-used for another review by: querying reviews that have the same topic and intention as the responded review, then identifying which part of the sentence that can replaced by a more local information to produce a new suitable response. 

After identifying the template, we are left with 60 templates, with an average of 1.6 placeholders per template. Table \ref{tab:placeholder} shows that \textbf{the placeholders include named entities (e.g. user name, app names), or hyperlinks (e.g. email, customer support link)}. While there is a high number of non-templated sentences (i.e. can be re-used as-is), these samples do not represent the actual distribution of these templated sentences in our data. For further studies, we would suggest the development of a technique to automatically identify templates.

\begin{table*}[t]
\small
  \centering
  \caption{Examples of response sentence templates}
    \begin{tabular}{p{25em}p{25em}}
    \toprule
    \textbf{Response sentence} & \textbf{Template}  \\
    \midrule
    any question o suggestion please contact us sending email to lbedeveloper@gmail.com we'll response asap  & any question o suggestion please contact us sending email to [email] we'll response asap  \\
    \midrule
    please tell us about your experience here https://goo.gl/forms/3pu3u1o2xg1frhsv2  & please tell us about your experience here [Link to a survey]  \\
    \midrule
    please create a ticket at our support website kefirgames.helpshift.com/a/last-day-on-earth we'll see how we can help you & please create a ticket at our support website [Link to Customer Support] we'll see how we can help you   \\
    \midrule
    please visit our faq at www.gotinder.com/faq for troubleshooting tips  & please visit our faq at [Link to a solution] for troubleshooting tips   \\
    \midrule
    if problem only with uc plz email us help@idc.ucweb.com & if problem only with [App Name] plz email us [email] \\
    \midrule
    Hi Denis, we're always looking to improve your experience with Uber  & Hi [User Name], we're always looking to improve your experience with [App Name] \\
    \midrule
    we're so sorry to hear about this vinit  & we're so sorry to hear about this [User Name] \\
    \midrule
    we appreciate you reaching out about the permissions zedge needs in order to work properly  & we appreciate you reaching out about the permissions [App Name] needs in order to work properly \\
    \bottomrule
    \end{tabular}%
  \label{tab:egTemplates}%
\end{table*}%

\subsubsection{Named entities}
As shown in Table \ref{tab:egTemplates}, some templates can contain Named-entities such as User Names, App Names. We observed that these entities are identifiable from the app information. It should be feasible to automatically infer them. 

\subsubsection{hyperlinks}
Hyperlinks are mostly to refer to an external source. There are three kinds of hyperlinks in our samples: (1) link to an external source of information; (2) link to a customer support email, or website; (3) link to a survey system.

The first type is usually used as a way to convey information that may have been too much to type in the response. For example, they could be a link to a solution, or to a forum, etc. Depending on how specific the description of the link is, its re-usability maybe vary. Examples are shown in Table \ref{tab:egTemplates}.

The second type, however, is just a way to refer to an external customer support system. Those systems are usually more comprehensive, and offer more resources to solve a customer's problem than the response can. From our observation, as long as the developer has an external customer support system, these responses with a customer support link can be re-used.

\begin{table}
  \centering
  \caption{Placeholder types in 100 random response samples}
    \begin{tabular}{lr}
    \toprule
       \textbf{Placeholder types} &  \textbf{Count} \\
    \hline
    Username &  25 \\
    Link &  21 \\
    Email &  15 \\
    App Name &  17 \\
    Company Name &  10 \\
    developers Name &  10 \\
    No place holder &  40 \\
    \bottomrule
    \end{tabular}%
  \label{tab:placeholder}%
\end{table}%

The last type is usually for developers to collect more detailed information on customer's experience that could not be collected from app reviews or within the app.

There could be more types of hyperlinks in responses, however, for the automated response system, we believe these three types are most important.

\subsection{What are the structures of full responses}

Toward generating a complete response, we also need to understand its structure. While observing the responses, we have noticed that responses often are composed of different parts in a specific order to answer to an intention in the reviews. To test this theory, we used the simple Hidden Markov Chain model~\cite{eddy1996hidden} to capture the intention chains if the reviews contain at least a major intention: a Request, or a Complaint. 

\begin{figure}
 \caption{Markov chain of response components for reviews containing Complaint intention}
	\includegraphics[width=0.45\textwidth]{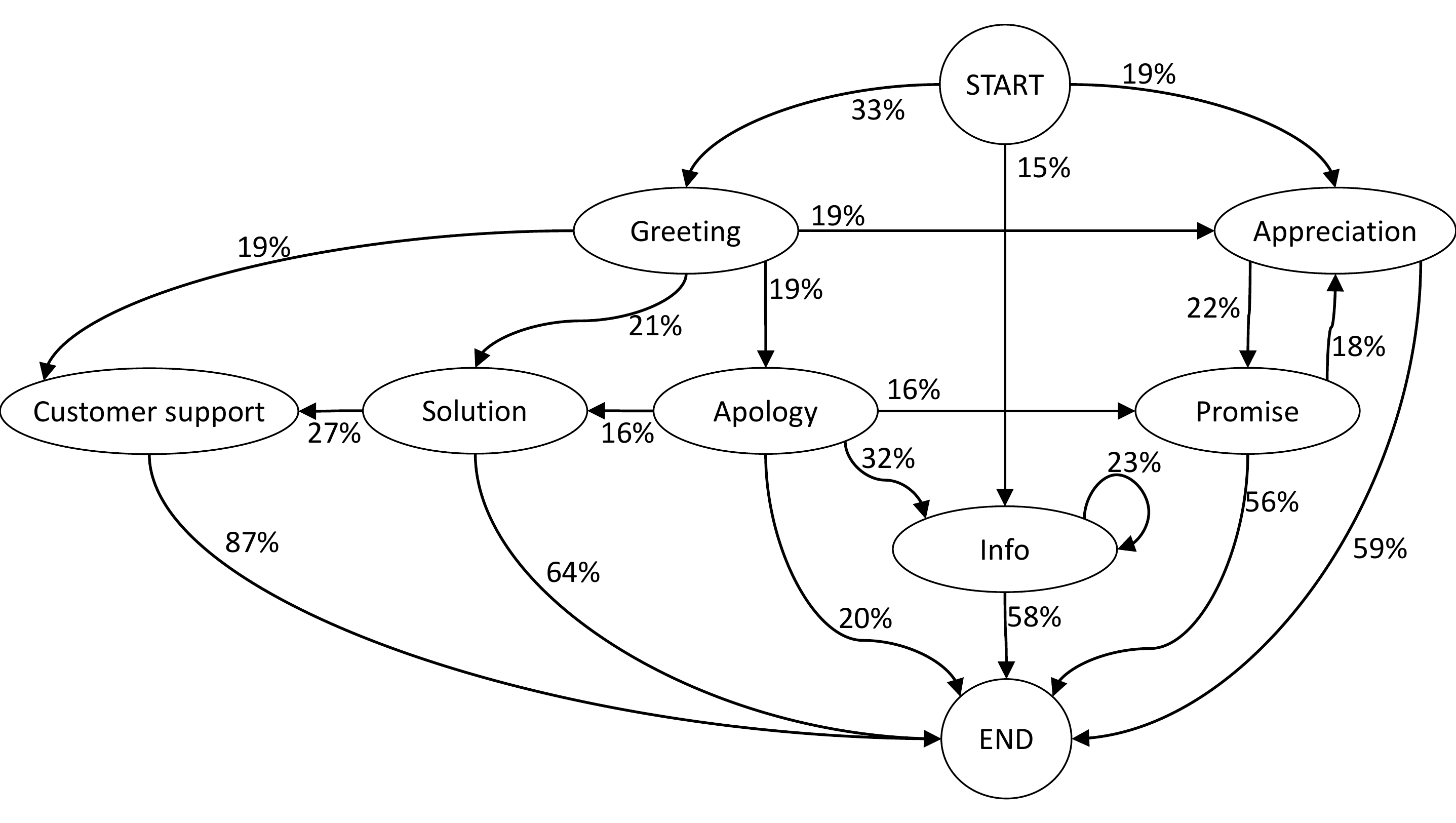}
 	\label{fig:markovComplaint}
\end{figure}

\begin{figure}
 \caption{Markov chain of response components for reviews containing Request intention}
	\includegraphics[width=0.45\textwidth]{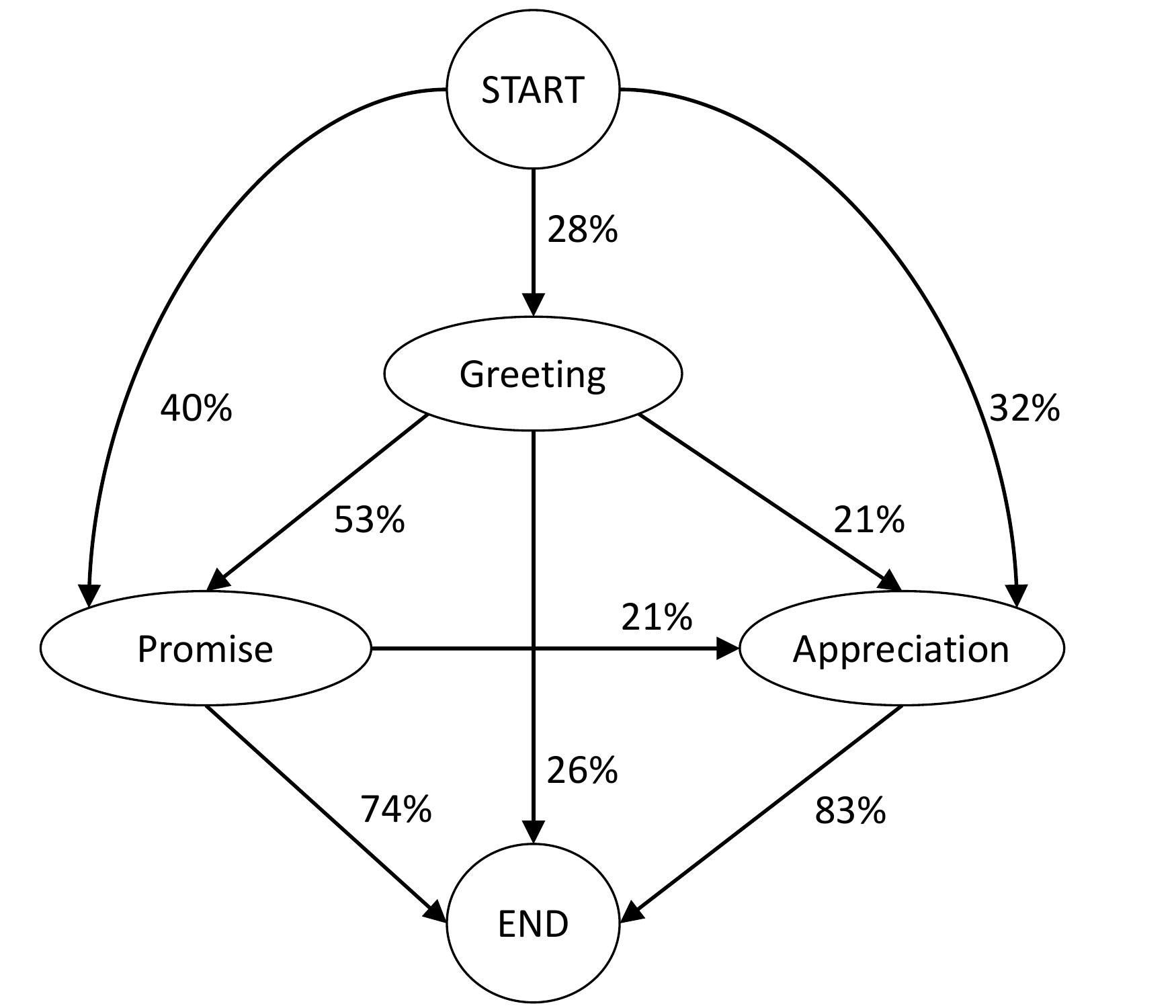}
 	\label{fig:markovRequest}
\end{figure}

Figure \ref{fig:markovComplaint} and \ref{fig:markovRequest} showed two different results for those aforementioned intentions. This s\textbf{uggests that building a Markov chain model for each review intention would be a feasible way to generate the components of a response.}

\section{Discussion}
\label{sec:discussion}
In this section, we discuss our findings and how they would be used to design an effective automated responding system for app reviews.

\textbf{Review rating is not a determining factor in prioritizing which review to respond to}. In prior work, the authors found that ratings for responded reviews are vary from app to app, but negative reviews have a higher likelihood of getting a response. Our results gave a similar conclusion, but we also found that positive ratings accounted for 36\% of our data, while negative ratings accounted for 50\% (the rest was neutral). This distribution was not a decisive win for the negative ratings, which means we need other metrics to determine which reviews should be responded.

\textbf{Some review topics (i.e. Feature/Functionality, App, Content, Update/version, GUI) are more prone to get a response than others.} The other topics (Improvement, Pricing, Resources, Security, Company, Model) was mentioned in less than 20\% of the reviews. While it is not conclusive that developers only focused on responding to reviews with the popular topics, because we did not know how they chose which topic to respond to, the skewed distribution of reviews on these topics may suggest that it is possible to rank the likelihood to be responded by topics. As Google Play Store and other stores may continue to hold a strict limit number of responses a day in the future, the automated system should prioritize the reviews with higher impact topics. Moreover, we realized that the data we have could not explain why developers chose such topics in the first place, we suggest future researches to solve this problem using one or both of the following approaches: (1) Set up an extensive interview with multiple developers to empirically study how they choose topics; (2) Allow the user of the Automated Responding System to rank the topics according to their preference. 

\textbf{Similar to topics, some intentions are more prone to get a response than others}. We found that more reviews with Complaint (47\%), Praise (39\%), Request (37\%) have a dominant presence in our dataset. A prior work \cite{DECA}, has already designed a method for determining intentions, which can be used to automatically classify intentions of sentences in reviews. However, this is similar to topics, we did not interview the developers to understand how they choose which intentions. Therefore, we recommend the same approaches: (1) Set up an extensive interview with multiple developers to empirically study how they choose intentions; (2) Allow the user of the Automated Responding System to rank the intentions according to their preference. 

\textbf{Responded reviews usually have at least 2 topics and intentions}. Hassan\etal has studied the likelihood of reviews getting responded for multiple quantitative metrics (e.g. text length, title length, sentiment, etc). In this study, we introduced two more metrics: the average number of topics and intentions. Our results suggest that developers may be more interested in responding to a review if it addresses multiple opinions. Therefore, these metrics should also be included when evaluating the features of the review ranking classification model in the automated system.

\textbf{Responses are composed of a different set of intentions.} When writing a response, the developers usually add various parts to it instead of just addressing the main points of the review. For example, Greetings, Farewells, and Apologies can be used to show courtesy to the customer, or soften their frustration. Promise, Solution, Information Giving, Information Seeking are usually to provide users with information that addresses their review. Sometimes, developers would redirect the users to an external Customer Support source instead. Each of those components has a usage, and we recommend the Automated System to use them as building blocks for a sensible response.

\textbf{81\% of response sentences can be re-used, however, different topics/apps/categories have different re-usability.} We found that the re-usability of past responses in a single app is varied by its review topics and intentions. Some topics (e.g. Security, Company, Improvement) and intentions (e.g. Praise, Information Giving) have a high likelihood of getting a re-used response. This implies that the Automated System can design a prediction model to infer what can be used to generate the response for each review. Moreover, we also observed that a past response can be reused for a different app if such a response describes the same feature in both apps. However, future studies are needed to find a method to recognize similar app features, as the current app store's categories may not necessarily imply the apps share those. For example, most Tool and Communication apps in our dataset cannot re-use each other's responses.

\textbf{Re-usable response can be templated}. We found that there are multiple types of templates from responses that can be reused in another context (i.e. different app/review) if we change their placeholders into a localized text (e.g. user name, app name, link). We recommend the Automated Responding System to create an automated approach to derive those templates, as they may require significant human labor to manually infer them.

\textbf{Markov Chain Model can be used to generate bare-components of responses}. We tested the Markov Chain model approach to capture the intention chains correlate to two major review intentions: Request and Complaint. The results shown in Figure \ref{fig:markovRequest} and \ref{fig:markovComplaint} implies that this is indeed feasible. However, this is just a pilot model, we recommend future researches to explore the generative models for all other intentions and different combinations of intentions and topics. 
\section{Threats to Validity}
\label{sec:validity}
\subsection{Construct validity}
In several of our quantitative studies, we chose a threshold for a maximum of 100 responded reviews to be analyzed as this number is reasonable for the scale of our manual effort. However, we are aware that this sample size may not represent the actual demographics of reviews and responses, as their content can be varied app by app, and developer by developer. Therefore, we encourage future researchers to further investigate in a larger scale study to verify and compare our findings.

Similarly, the minimum threshold of analyzed reviews (i.e. 50 reviews) also left out several topics and intention that can contribute to the understanding of how they would affect responses. However, the given the total amount of reviews (2,274), the left out topics/intentions each contributed about 2.1\%, which may not be represented in our data. To study the effects of them to the re-usability of responses, we plan to collect more data in the future.

\subsection{Internal validity}

In this study, we collected and analyzed the top trending apps in 2019 for a three months period. We are aware that this data may not be enough to provide a complete insight into the review responses of the entire app store. We have asked authors of the prior works~\cite{hassan2017reviewDialog} for their dataset, which is much larger than ours. Regretfully, under the restriction in sharing data from Google, they were not able to share it with us. We suggest that future works should collect more apps in a longer period for a more complete overview of the app ecosystem. 

Our studies are largely dependent on the accuracy of the labeling process from our authors. It is possible that there are errors in our annotations. To reduce the margin of error, only two authors with at least 4 years experience in working with mobile app reviews were chosen to annotate. Moreover, we ensure the discussion step is done responsibly as the annotators had to explain their implications for each label. However, since the authors have to agree with each label, there is a chance that they are still biased by our own experience and intuition despite the reasonable effort.

\subsection{External validity}

Similar to the external validity that was discussed by Hassan\etal, we also have done the same crawling process to minimize the losses of reviews in our dataset. Moreover, we removed all paid apps from the study as we also agreed that the pricing can be a major factor to determine how the developer would respond to reviews. However, since how that pricing would affect the response was not covered in our study, we encourage future researchers to tackle this question.

\section{Related works}
\label{sec:related}
There is a number of empirical and exploratory studies on the importance of app's reviews in the app development process. In \cite{vasa2012preliminary}, Vasa\etal made an exploratory study about how users input their reviews on app stores and what could affect the way they write reviews. Later,  Hoon\etal \cite{hoon2013analysisReviewLandscape} analyzed nearly 8 million reviews on Apple AppStore to discover several statistical characteristics to suggest developers constantly watching for the changes in user's expectations to adapt their apps. Again on Apple App Store, an empirical study about user's feedback was made by Pagano \etal\cite{feedbackEmpirical}. Similarly, Khalid \etal suggests that there are at least 12 types of complaints about iOS apps~\cite{khalid2013identifying}. They explored various aspects that influent user reviews such as time of release, topics and several properties including quality and constructiveness to understand their impacts on apps. Nayebi\etal~\cite{nayebi2018app} studied the correlation between app reviews and tweets. Palomba\etal had conducted a study \cite{palomba2018crowdsourcing} of whether user reviews really are taken into account by developers in app development. Scoccia\etal~\cite{scoccia2018investigation} analyzed 3,754 permission related reviews and found that permission related issues are widespread even in new apps. Truelove\etal~\cite{truelove} studied 90,000 reviews of IoT devices apps to identify related issues.

Some other works resulted in complete toolset or prototypes such as Wiscom \cite{Wiscom} or MARA\cite{Iacob:2013:MARA}, or AR-Miner \cite{Chen:2014:AR-miner}. Chen et al. propose a computational framework to extract and rank informative reviews at the sentence level. They adopt the semi-supervised algorithm Expectation Maximization for Naive Bayes (EMNB)\cite{nigam1999using} to classify between informative and non-informative reviews. To rank the reviews, they use a ranking schema based on the meta-data of reviews and suggest the most informative ones. Gao\etal also provided INFAR\cite{gao2018infar} to analyze app reviews from multiple analysis dimensions. Later on, MARK \cite{vu2015MARKtool} proposed a keyword-based approach to discovering topics and trends using their semantic meaning. After that, Panichella\etal~\cite{panichella2015can} created the SURF tool that identifies the important opinions from user reviews by general topics and several intentions developed by Iacob\etal 

However, to the best of our knowledge, there is currently no automated tool to support developers to respond to app reviews. This was either because of the use of canned-response tools~\cite{esplin2015mechanism} is sufficient, or because the technical difficulties of app responses were not properly studied.

Responses of app reviews have been studied several times in recent years. Oh\etal (2013)~\cite{oh2013facilitating} surveyed 100 users and reported that the users prefer to post reviews on the app store to an outside source. McIlroy\etal (2015)~\cite{mcilroy2017worth} published a study about how ratings are affected by developers' response. They found that users often increase their rating after receiving a response from the developers. Hassan\etal~\cite{hassan2017reviewDialog} analyzed the dialogues between developers and users, and found several findings that would motivate the creation of an Automated Responding System for reviews. Kendall\etal(2019)\cite{bailey2019examining}, have further investigated in the Feedback-loops created by developers and users and found that several topics would likely to trigger such loops in the review system. Lastly, Noel\etal (2019)~\cite{noei2019too} studied the priorities of reviews that would need developers to look into and also found that those priorities correlated to several important topics.

\section{Conclusion}
\label{sec:conclusion}
The recent researches have motivated for the creation of an Automated Responding System for app reviews. However, there were several questions about the characteristics of reviews that needed to be studied before the realization of such a system. 

In this study, we crawled approximately 650 thousand reviews from Google Play Store, then manually labeled 3,212 pairs of review and response with 18 thousand sentences. We then further analyzed and observe this data from the perspective of building an Automated Responding System. Some of our most notable findings are:
\begin{itemize}
    \item Rating, review topics, review intentions likely to have an impact on the priority of reviews to be responded.
    \item Responses are built by multiple components, each has a different intention.
    \item 81\% of response sentences can be re-used in a new review, however, the re-usability is varied by different topics/app/categories. The responses do not necessarily to be of the same old text, certain parts of them can be placeholders (i.e. responses can become templates).
    \item Markov Chain Model is a feasible way to generate response components.
\end{itemize}

We explained how we conducted each study, and then discussed the implications of our findings toward building an Automated Responding system for app reviews.

\bibliographystyle{IEEEtran}
\bibliography{IEEEabrv,automatedResponse}

\begin{thebibliography}{10}
\providecommand{\url}[1]{#1}
\csname url@samestyle\endcsname
\providecommand{\newblock}{\relax}
\providecommand{\bibinfo}[2]{#2}
\providecommand{\BIBentrySTDinterwordspacing}{\spaceskip=0pt\relax}
\providecommand{\BIBentryALTinterwordstretchfactor}{4}
\providecommand{\BIBentryALTinterwordspacing}{\spaceskip=\fontdimen2\font plus
\BIBentryALTinterwordstretchfactor\fontdimen3\font minus
  \fontdimen4\font\relax}
\providecommand{\BIBforeignlanguage}[2]{{%
\expandafter\ifx\csname l@#1\endcsname\relax
\typeout{** WARNING: IEEEtran.bst: No hyphenation pattern has been}%
\typeout{** loaded for the language `#1'. Using the pattern for}%
\typeout{** the default language instead.}%
\else
\language=\csname l@#1\endcsname
\fi
#2}}
\providecommand{\BIBdecl}{\relax}
\BIBdecl

\bibitem{mcilroy2017worth}
S.~McIlroy, W.~Shang, N.~Ali, and A.~E. Hassan, ``Is it worth responding to
  reviews? studying the top free apps in google play,'' \emph{IEEE Software},
  vol.~34, no.~3, pp. 64--71, 2017.

\bibitem{vu2015MARK}
P.~M. Vu, T.~T. Nguyen, H.~V. Pham, and T.~T. Nguyen, ``Mining user opinions in
  mobile app reviews: A keyword-based approach (t),'' in \emph{ASE}, 2015.

\bibitem{hassan2017reviewDialog}
S.~Hassan, C.~Tantithamthavorn, C.-P. Bezemer, and A.~E. Hassan, ``Studying the
  dialogue between users and developers of free apps in the google play
  store,'' \emph{Empirical Software Engineering}, pp. 1--38, 2017.

\bibitem{surf}
A.~Di~Sorbo, S.~Panichella, C.~V. Alexandru, J.~Shimagaki, C.~A. Visaggio,
  G.~Canfora, and H.~C. Gall, ``What would users change in my app? summarizing
  app reviews for recommending software changes,'' in \emph{FSE}, 2016.

\bibitem{reviewresponse}
P.~M. Vu, T.~T. Nguyen, and T.~T. Nguyen, ``Why do app reviews get responded? a
  preliminary study of the relationship between reviews and responses in mobile
  apps,'' in \emph{ACMSE}.\hskip 1em plus 0.5em minus 0.4em\relax ACM, 2019.

\bibitem{alpaca}
P.~M. Vu, T.~T. Nguyen, H.~V. Pham \emph{et~al.}, ``Alpaca-advanced linguistic
  pattern and concept analysis framework for software engineering corpora,'' in
  \emph{Proceedings of the 40th International Conference on Software
  Engineering: Companion Proceeedings}.\hskip 1em plus 0.5em minus 0.4em\relax
  ACM, 2018, pp. 284--285.

\bibitem{strauss1997grounded}
A.~Strauss and J.~M. Corbin, \emph{Grounded theory in practice}.\hskip 1em plus
  0.5em minus 0.4em\relax Sage, 1997.

\bibitem{khandkar2009open}
S.~H. Khandkar, ``Open coding,'' \emph{University of Calgary}, vol.~23, p.
  2009, 2009.

\bibitem{fleiss1973equivalence}
J.~L. Fleiss and J.~Cohen, ``The equivalence of weighted kappa and the
  intraclass correlation coefficient as measures of reliability,''
  \emph{Educational and psychological measurement}, vol.~33, no.~3, pp.
  613--619, 1973.

\bibitem{Chen:2014:AR-miner}
N.~Chen, J.~Lin, S.~C. Hoi, X.~Xiao, and B.~Zhang, ``Ar-miner: mining
  informative reviews for developers from mobile app marketplace,'' in
  \emph{ICSE}, 2014.

\bibitem{eddy1996hidden}
S.~R. Eddy, ``Hidden markov models,'' \emph{Current opinion in structural
  biology}, vol.~6, no.~3, pp. 361--365, 1996.

\bibitem{DECA}
A.~Di~Sorbo, S.~Panichella, C.~A. Visaggio, M.~Di~Penta, G.~Canfora, and
  H.~Gall, ``Deca: development emails content analyzer,'' in \emph{ICSE}, 2016.

\bibitem{vasa2012preliminary}
R.~Vasa, L.~Hoon, K.~Mouzakis, and A.~Noguchi, ``A preliminary analysis of
  mobile app user reviews,'' in \emph{OzCHI}, 2012.

\bibitem{hoon2013analysisReviewLandscape}
L.~Hoon, R.~Vasa, J.~Schneider, and J.~Grundy, ``An analysis of the mobile app
  review landscape: Trends and implications,'' Tech. Rep., 2013.

\bibitem{feedbackEmpirical}
D.~Pagano and W.~Maalej, ``User feedback in the appstore: An empirical study,''
  in \emph{RE}, 2013.

\bibitem{khalid2013identifying}
H.~Khalid, ``On identifying user complaints of ios apps,'' in \emph{Proceedings
  of the 2013 International Conference on Software Engineering}.\hskip 1em plus
  0.5em minus 0.4em\relax IEEE Press, 2013, pp. 1474--1476.

\bibitem{nayebi2018app}
M.~Nayebi, H.~Cho, and G.~Ruhe, ``App store mining is not enough for app
  improvement,'' \emph{Empirical Software Engineering}, pp. 1--31, 2018.

\bibitem{palomba2018crowdsourcing}
F.~Palomba, M.~Linares-V{\'a}squez, G.~Bavota, R.~Oliveto, M.~Di~Penta,
  D.~Poshyvanyk, and A.~De~Lucia, ``Crowdsourcing user reviews to support the
  evolution of mobile apps,'' \emph{Journal of Systems and Software}, vol. 137,
  pp. 143--162, 2018.

\bibitem{scoccia2018investigation}
G.~L. Scoccia, S.~Ruberto, I.~Malavolta, M.~Autili, and P.~Inverardi, ``An
  investigation into android run-time permissions from the end users'
  perspective,'' in \emph{Proceedings of the 5th International Conference on
  Mobile Software Engineering and Systems}.\hskip 1em plus 0.5em minus
  0.4em\relax ACM, 2018, pp. 45--55.

\bibitem{truelove}
A.~Truelove, F.~N.~Chowdhury, O.~Gnawali, and M.~A.~Alipour, ``Topics of
  concern: Identifying user issues in reviews of iot apps and devices,'' in
  \emph{International Workshop on Software Engineering Research \& Practices
  for the Internet of Things}, 2019.

\bibitem{Wiscom}
B.~Fu, J.~Lin, L.~Li, C.~Faloutsos, J.~Hong, and N.~Sadeh, ``Why people hate
  your app: Making sense of user feedback in a mobile app store,'' ser. KDD
  '13, 2013.

\bibitem{Iacob:2013:MARA}
C.~Iacob and R.~Harrison, ``Retrieving and analyzing mobile apps feature
  requests from online reviews,'' in \emph{MSR}, 2013.

\bibitem{nigam1999using}
K.~Nigam, J.~Lafferty, and A.~McCallum, ``Using maximum entropy for text
  classification,'' in \emph{IJCAI-99 workshop on machine learning for
  information filtering}, vol.~1, 1999, pp. 61--67.

\bibitem{gao2018infar}
C.~Gao, J.~Zeng, D.~Lo, C.-Y. Lin, M.~R. Lyu, and I.~King, ``Infar: insight
  extraction from app reviews,'' in \emph{Proceedings of the 2018 26th ACM
  Joint Meeting on European Software Engineering Conference and Symposium on
  the Foundations of Software Engineering}.\hskip 1em plus 0.5em minus
  0.4em\relax ACM, 2018, pp. 904--907.

\bibitem{vu2015MARKtool}
P.~M. Vu, H.~V. Pham, T.~T. Nguyen, and T.~T. Nguyen, ``Tool support for
  analyzing mobile app reviews,'' in \emph{ASE}, 2015.

\bibitem{panichella2015can}
S.~Panichella, A.~Di~Sorbo, E.~Guzman, C.~A. Visaggio, G.~Canfora, and H.~C.
  Gall, ``How can i improve my app? classifying user reviews for software
  maintenance and evolution,'' in \emph{ICSME}, 2015.

\bibitem{esplin2015mechanism}
A.~Esplin, K.~Whitney, N.~DiJulio, J.~Oh, I.~Baek, and I.~Bush, ``Mechanism for
  facilitating dynamic generation and transmission of canned responses on
  computing devices,'' Jul.~2 2015, uS Patent App. 14/140,641.

\bibitem{oh2013facilitating}
J.~Oh, D.~Kim, U.~Lee, J.-G. Lee, and J.~Song, ``Facilitating developer-user
  interactions with mobile app review digests,'' in \emph{CHI'13 Extended
  Abstracts on Human Factors in Computing Systems}.\hskip 1em plus 0.5em minus
  0.4em\relax ACM, 2013, pp. 1809--1814.

\bibitem{bailey2019examining}
K.~Bailey, M.~Nagappan, and D.~Dig, ``Examining user-developer feedback loops
  in the ios app store,'' in \emph{Proceedings of the 52nd Hawaii International
  Conference on System Sciences}, 2019.

\bibitem{noei2019too}
E.~Noei, F.~Zhang, and Y.~Zou, ``Too many user-reviews, what should app
  developers look at first?'' \emph{IEEE Transactions on Software Engineering},
  2019.

\end{thebibliography}

\end{document}